\documentclass{kluwer}    

\newdisplay{guess}{Conjecture}
\usepackage[dvips]{graphicx}
\begin{document}                                                                                   
\begin{article}
\begin{opening}         
\title{Stellar Populations in Bulges of Spiral Galaxies} 
\author{Pascale \surname{Jablonka}}  
\runningauthor{Pascale Jablonka}
\runningtitle{Stellar Populations in Bulges of Spiral Galaxies}
\institute{D.A.E.C., UMR 8631, Place Jules Janssen, 92195 MEUDON, FRANCE}
\author{Javier \surname{Gorgas}}
\institute{Dpt. de Astrofisica, Universidad Complutense, Madrid, SPAIN}
\author{Paul \surname{Goudfrooij}}
\institute{Space Telescope Science Institute, Baltimore, USA}
\date{December 1, 2001}

\begin{abstract}
We present a few results of a work in progress tackling the
radial spectroscopic properties of bulges of spiral galaxies.
\end{abstract}
\keywords{Spiral galaxies, Evolution of galaxies, Stellar population }
\end{opening}           

\section{Introduction}  

We have conducted a spectroscopic analysis of the bulges of a sample
of 31 spiral galaxies, spanning the Hubble sequence from S0 to Scd
type, in order to study the radial distribution of their stellar
population properties. This is the first sample of this size, allowing
at last some statistics. We have in mind the following considerations :

\begin{itemize}

\item Radial spectral information brings insights on the structure of the
systems which are considered, as it unveils the light/mass spatial
distribution.  It also gives important clues on time scales at play:
either directly from age indicators, or, particularly for old
populations, by considering abundance ratios of chemical elements
appearing with the explosion of different supernova types, like type Ia
and type II.

\item Moreover, debates on the formation and evolution of galaxies are,
specially for nearby galaxies, more than often based on observations
of {\it central} indices, or, more precisely, integrated indices in
apertures centered on the galaxy nuclei and of small sizes when
compared to the size of the galaxies.  Meanwhile, models generally
follow/predict the global properties of the galaxies, which would
find a more natural echo in {\it mean} observational
quantities. This is also the case of observations at high redshift.
Tracking the spatial distribution of the stellar population allows one
to trace back these galactic {\it mean} spectral indices.

\item Numerical simulations nowadays offer a good deal of various proposals
for the origin of the formation of bulges, however data are crucially
missing to tackle crucial questions like : how far can we push
the parallel between ellipticals and bulges ? by how much do the
discs and bulges interplay and when in the galaxy history ?

\end{itemize}

\section{Observations and results}

We have selected purely edge-on spirals in order to free our spectra
from any disc light. We also centered the slit of the spectrograph and
placed it perpendicular to the plane of the discs to avoid any trend
due to the rotation of the bulges. We finally could measure
our indices up to at least one bulge effective radius ($r_{\rm eff}$)
and upx to 5$r_{\rm eff}$. 

We chose to measure our spectral features in the Lick system (Worthey
et al. 1994), for purposes of comparison with large samples of
ellipticals.  Our interest for the present time is not to give
absolute values in age or metallicity (Z), but to understand whether
variations of our indices are due to a radial change in age, in
metallicity or in both.

\begin{figure}[h]
\centerline{\includegraphics[width=12pc]{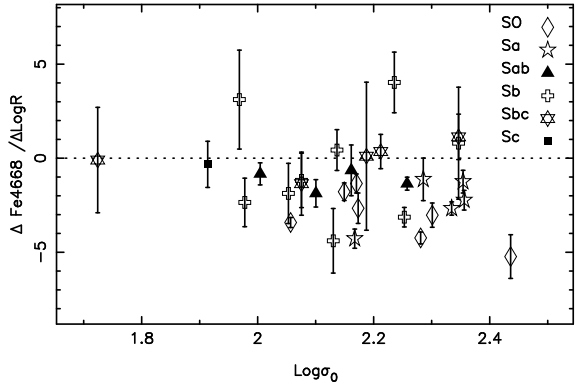}}
\caption[]{Radial gradient of the Fe4668 index as a function of
the bulge central velocity dispersion.}
\label{Fe4668}
\end{figure}

As there is no one-to-one relation between any individual Lick index
and age or even the abundance of one single chemical element (Worthey
1994), we use as many indices as possible and consider their different
relative sensitivities to age and metallicity. The wavelength coverage
of our spectra enable us to measure 25 indices, from H$\delta$ to TiO.

We present here the example of Fe4668 and H$\gamma_A$.  Fe4668 is
around 5 times more sensitive to metallicity than to the logarithm of
the age, while H$\gamma_A$ is about as sensitive to Log(age) as to
metallicity. Figure~1 and Figure~2 present the radial gradients of
these two indices as a function of the bulge central velocity
dispersion. A different symbol is used for each Hubble type. The error
bars of our measurements are shown, they are the result of a detailed
error propagation from the first steps of data reduction to the final
measurements. We show as dotted line the position of a gradient slope
of 0, which would correspond to the absence of any radial variation of
the index.

The distribution of $\Delta{\rm Fe4668}/\Delta\log r$ is flat with a
mean value of $-2.5$ and a r.m.s.  dispersion of 1.4. The distribution
of $\Delta{\rm H}_{\gamma_A}/\Delta\log r$ is flat with a mean value
of 1.5 and a dispersion of 0.8.

\begin{figure} [h]
\centerline{\includegraphics[width=12pc]{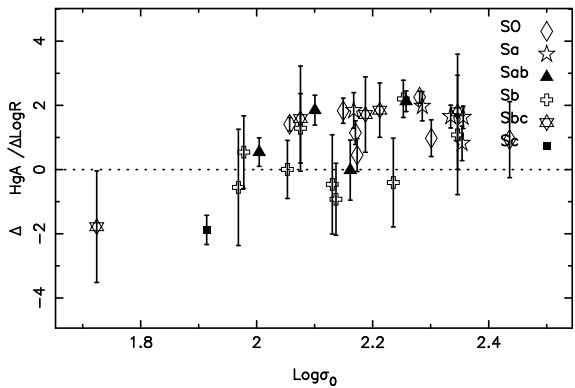}}
\caption[]{Radial gradient of the H$\gamma_A$ index as a function of
the bulge central velocity dispersion.}
\label{HgammaA}
\end{figure}

Two remarks can be made : (Note : We discuss the mean observed trends
and do not address any individual galaxy that might deviate slightly
from those trends)

\begin{itemize}

\item While Fe4668 decreases, H$\gamma_A$ increases from the center to
the outer edges of the bulges. The clear radial change of Fe4668, both
by itself and as compared to the variation of H$\gamma_A$, on one
hand, the sign of the slope of the gradients, on the other hand,
support the view of a variation in metallicity as primary underlying
factor, the bulge inner parts being more metal-rich than the outer
regions.  This is : a pure age gradient would make Fe4668 slope about
ten times smaller than that observed, and $\Delta{\rm
H}_{\gamma_A}/\Delta\log r$ greater than $\Delta{\rm
Fe4668}/\Delta\log r$, therefore this hypothesis is discarded.

\item Neither H$\gamma_A$ nor Fe4668 radial gradients show any
dependence on the Hubble type of the galaxies. This is also the case
for the other indices. The loose connection between the
morphological classification of the spirals and the bulge stellar
population had been previously noticed for the central indices of
bulges (Jablonka, Arimoto \& Martin 1996). But while the bulge central
indices are found to be related to the bulge central velocity
dispersion, there is no comparable trend for the radial gradients
(except for Mg$_2$).

\end{itemize}

As to the comparison between bulges and ellipticals, interestingly,
the relation between gradients in ellipticals and their central
velocity dispersion is a longstanding debate (Gonzalez \& Gorgas
1996; Kobayashi \& Arimoto 1999). A further comparison with the
results quoted in Gorgas et al. (1997) reveals that the mean slope of
the Mg$_2$ gradient in ellipticals coincides with the values we find
for the bulges of our sample ($-0.055$ $\pm$ 0.025 vs $-0.047$ $\pm$
0.023)

\section{Conclusion}

While by no means, we have exhausted all the information of
our dataset yet, we already see quite clearly that any scenario of formation
of bulges requiring long time scales would be in contradiction 
with the observations. We also stress the apparent continuity
between elliptical and bulge properties.

\end{article}
\end{document}